\documentclass[3p,times,procedia]{elsarticle}
\usepackage{nupha_ecrc}
\volume{00}
\firstpage{1}
\journalname{Nuclear Physics A}
\runauth{R.~Soltz}
\jid{nupha}
\jnltitlelogo{Nuclear Physics A}
\usepackage{amssymb}

\begin{document}

\begin{frontmatter}

\dochead{XXVIIIth International Conference on Ultrarelativistic Nucleus-Nucleus Collisions\\ (Quark Matter 2019)}
\title{A Comprehensive Monte Carlo Framework for Jet-Quenching}
\author{Ron Soltz}
\address{Wayne State University and Lawrence Livermore National Laboratory}
\ead{soltz1@llnl.gov}

\begin{abstract}
This article presents the motivation for developing a comprehensive modeling framework in which different models and parameter inputs can be compared and evaluated for a large range of jet-quenching observables measured in relativistic heavy-ion collisions at RHIC and the LHC.  The concept of a framework us discussed within the context of recent efforts by the JET Collaboration, the authors of JEWEL, and the JETSCAPE collaborations.  The framework ingredients for each of these approaches is presented with a sample of important results from each.  The role of advanced statistical tools in comparing models to data is also discussed, along with the need for a more detailed accounting of correlated errors in experimental results.
\end{abstract}


\end{frontmatter}


\section{Framework Motivation and Ingredients}
\label{intro}

The benefit of using a mutli-stage model with an advanced statistics package to study the properties and evolution of the quark-gluon plasma created in heavy ion collisions has been demonstrated in the soft physics sector by the work of the MADAI collaboration~\cite{Novak:2013tf} and the Duke QCD theory group~\cite{Bernhard:2016wl}.  It stands to reason that hard physics sector may benefit from a similar approach.  Whereas the methods for calculating soft observables have converged around viscous hydrodynamics preceded by an initial glasma+hydrodynamization and followed by Cooer-Frye hadronization into to a hadronic cascade, the methodology for jet-quenching is more varied, and may itself require a multi-stage approach~\cite{Majumder:2015wy}.  Thus, for hard sector, a framework is needed to provide the flexibility of testing the inherent assumptions within corresponding models, and for adding, removing, or modifying different stages or mechanisms of energy loss.

\begin{figure}[h]
\centering
\includegraphics[width=0.8\linewidth]{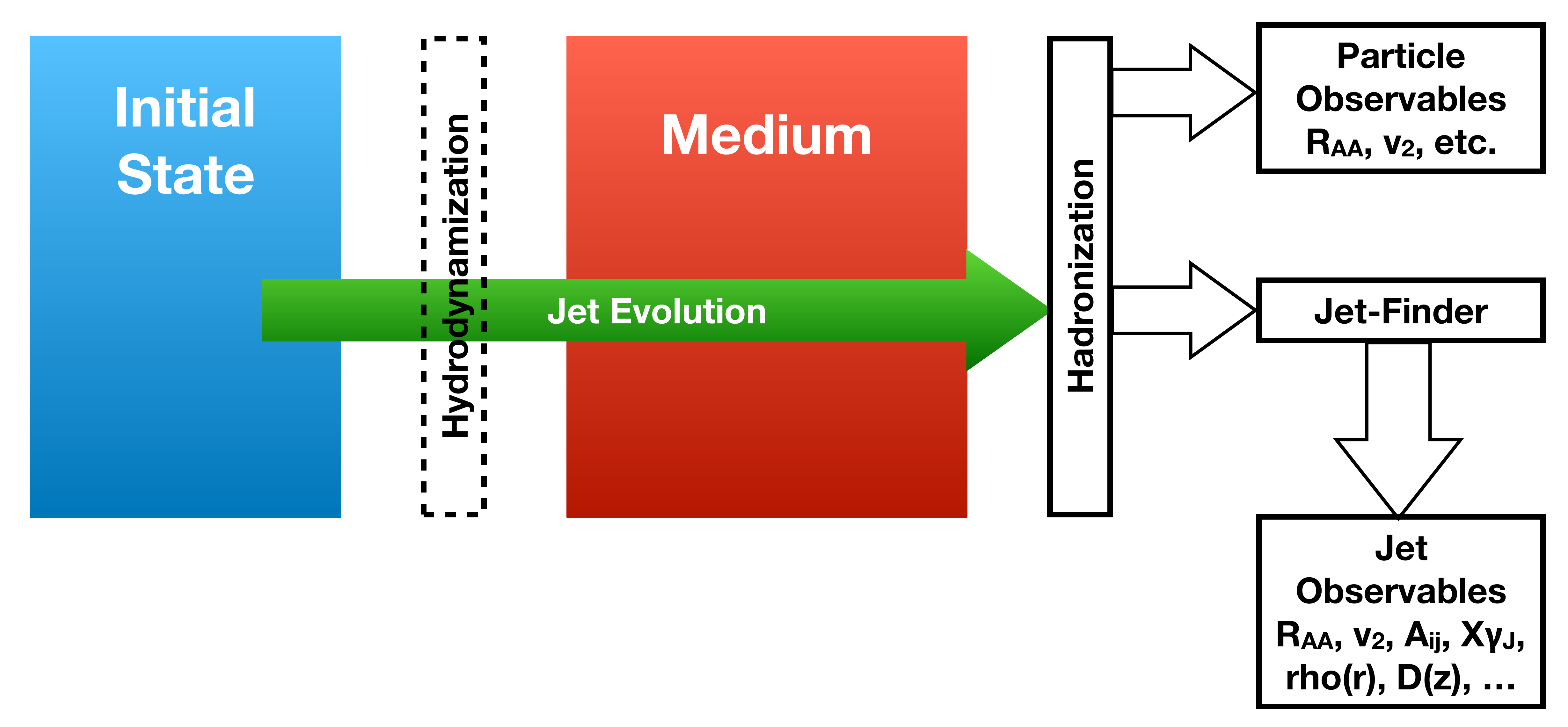}
\caption{Essential ingredients for a jet-quenching framework.}
\label{fig_framework_ingredients}
\end{figure}

Figure~\ref{fig_framework_ingredients} shows the essential ingredients for such a framework.  It includes the elements currently used to calculate soft observables and adds in components for jet evolution and a jet-finder to calculate jet observables.  In the next section we will discuss how the work of the JET Collaboration, and the JEWEL, and JETSCAPE computational approaches map onto this framework.

\section{Framework Examples}
\label{examples}

\subsection{JET Collaboration}
\label{jet}

The first steps towards building a true framework for jet-quenching were taken by the JET Collaboration.  Their goal was to develop a consistent approach for determining $\hat{q}=\langle(\Delta k_T)^2\rangle / L$, the energy-loss metric for measuring the transverse diffusion per unit length for a jet or parton traversing a dense nuclear-medium~\cite{Burke:2014ja}.  In this work, five different approaches were used to constrain $\hat{q}$ with the leading hadron nuclear modification, $R_{AA}$.  All approaches used a 2D+1 viscous hydrodynamic profile (except for HT-BW which used 3+1D ideal hydro) with central Glauber initial conditions (except for HT-M which used MC-KLN for initial conditions).  

\begin{figure}[h]
\centering
\includegraphics[width=0.5\linewidth]{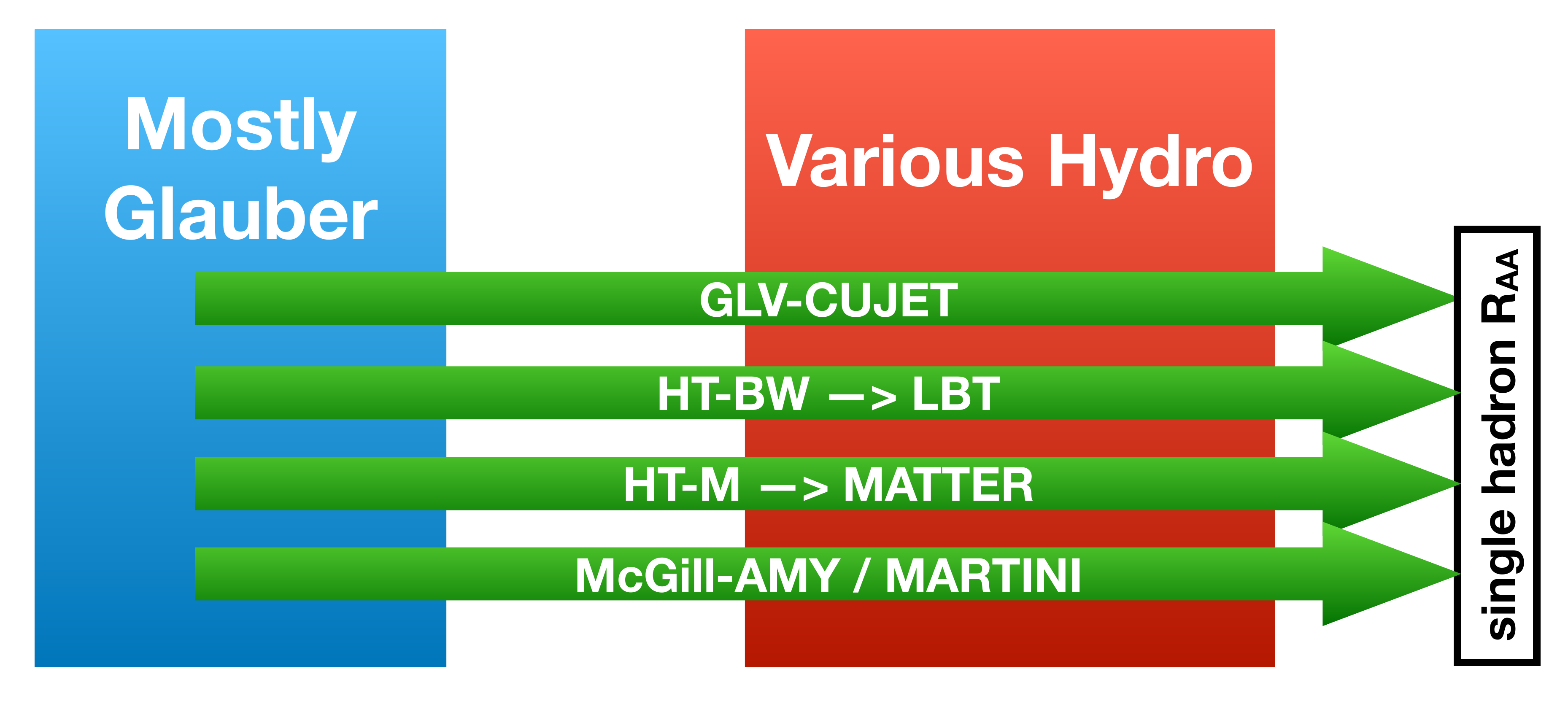}
\includegraphics[width=0.3\linewidth]{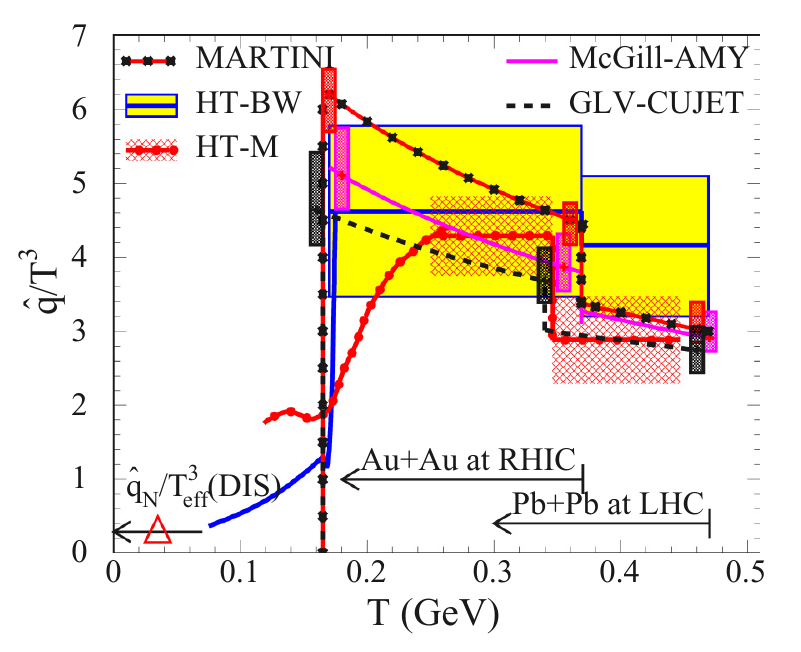}
\caption{(Left) JET collaboration approach as a jet-quenching framework and (right) $\hat{q}$ evaluations obtained using this approach.}
\label{fig_jet}
\end{figure}

Parameterizations for $\hat{q}/T^3$ were allow to differ for central collisions at RHIC and the LHC, but were otherwise independent of temperature during the evolution of a single collision system.  Chi-squared minimizations were used to determine optimal values for the leading hardon nuclear-monification factor, $R_{AA}$.  The left panel of Figure~\ref{fig_jet} shows the general framework schematic, and the right panel shows the corresponding values of $\hat{q}/T^3$ for each approach.  Although theses comparisons were not performed within single framework, the ability to compare models and data under a consistent set of assumptions represents a significant step in this direction.

\subsection{JEWEL}
\label{jewel}

\begin{figure}[h]
\centering
\includegraphics[width=0.8\linewidth]{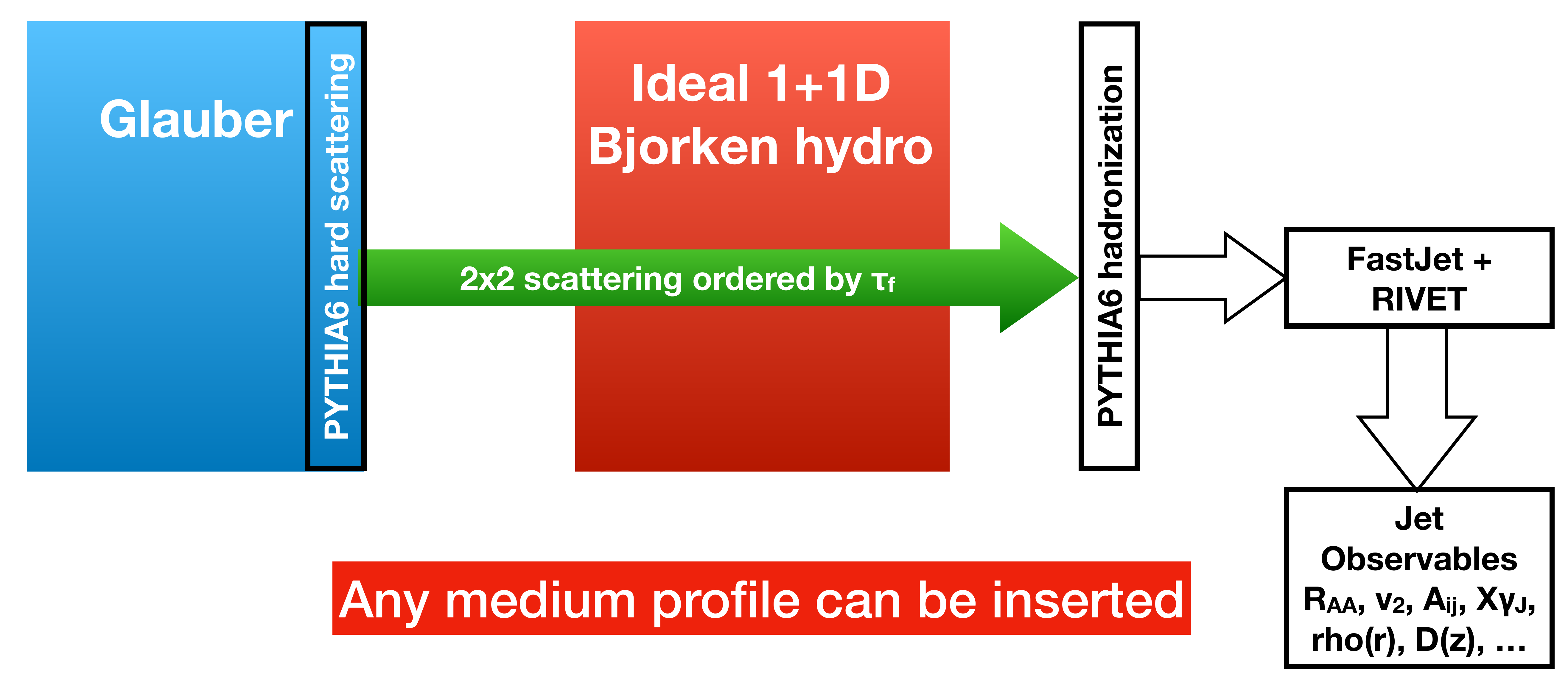}
\caption{JEWEL representation as jet-quenching framework.}
\label{fig_jewel}
\end{figure}

The first genuine Monte Carlo jet-quenching framework to match the criteria shown in Figure~\ref{fig_framework_ingredients} is JEWEL~\cite{Zapp:2011gy,Zapp:2012ja,Zapp:2013hh,Elayavalli:2016bc,Elayavalli:2017eb}, which includes both collisional and radiative energy-loss with formation time-ordering.  This framework can be implemented with any medium description specifying the space-time and momentum distribution of scattering centers.  The current implementation of JEWEL uses PYTHIA6~\cite{Sjostrand:2006aa} for both the initial hard-scattering and final hadronization.  Jet observables are calculated in FastJet~\cite{Cacciari:2012br} and RIVET~\cite{bierlich:2020aa} for comparison to experimental, and the medium description is calculated with an ideal 1+1D Bjorken hydrodynamics with Glauber initial conditions.  Figure~\ref{fig_jewel} shows this implementation of the JEWEL framework.

\begin{figure}[h]
\centering
\includegraphics[width=0.99\linewidth]{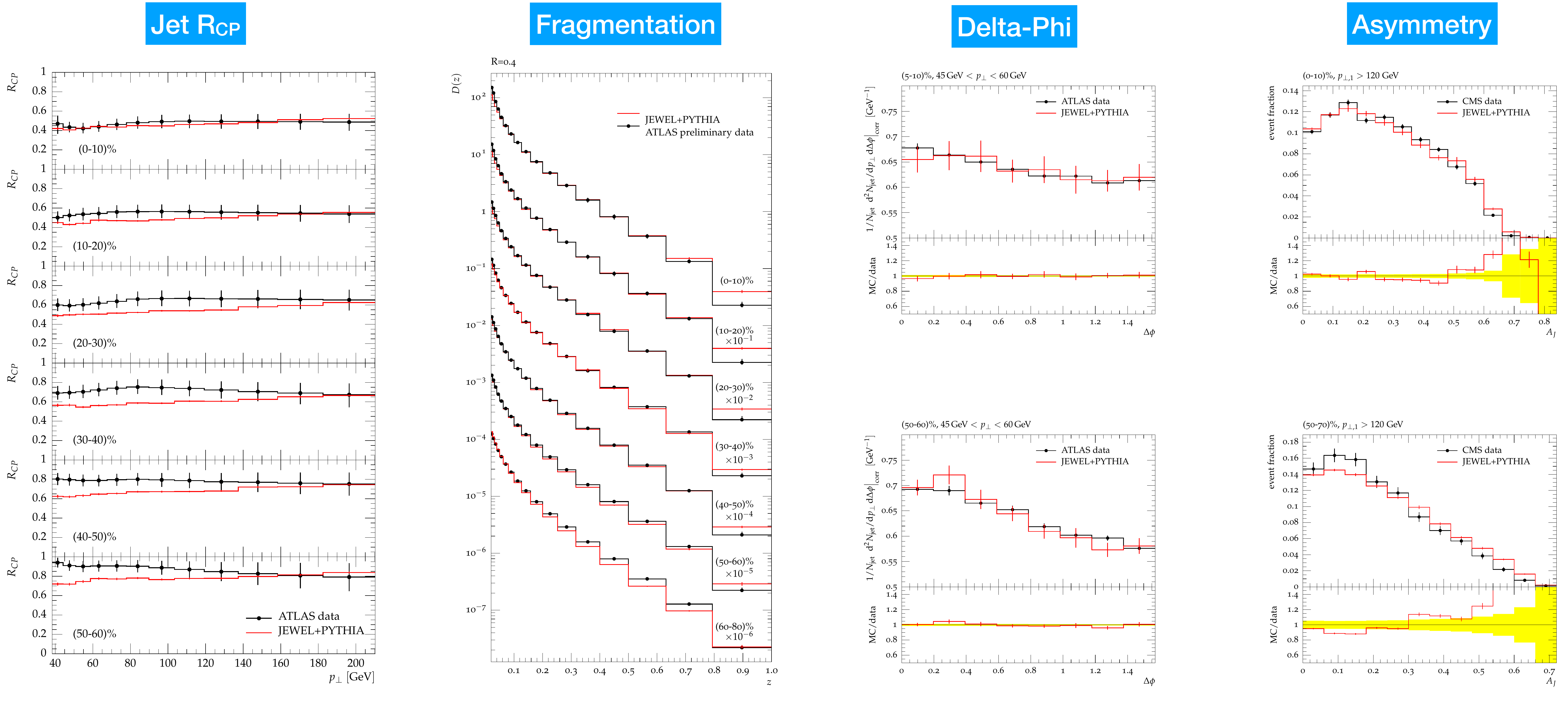}
\caption{JEWEL results~\cite{Zapp:2013hh} for Jet nuclear modification, fragmentation, reaction-plane angle, and asymmetry for Pb+Pb at $\sqrt{s_{NN}}$=2.76 TeV compared to measurements from ATLAS~\cite{Aad:2013hx,Aad:2013dq,ATLASCollaboration:2014fz} and CMS~\cite{CMSCollaboration:2012aa}}
\label{fig_jewel_xjg}
\end{figure}

A large set jet observables for Pb+Pb at $\sqrt{s_{NN}}$=2.76 TeV calculated within JEWEL and compared to ATLAS measurements are shown in Figure~\ref{fig_jewel_xjg}.
Most observables show reasonable agreement except for the $R_{CP}$ nuclear modification factor in the peripheral region, which is problematic for many models.  The ratio of MC to data shows that the JEWEL results are within the systematic error band (yellow bars) for the asymmetry.  
The role of systematic errors in model-to-data comparisons will be discussed further in Section~\ref{errors}.

\subsection{JETSCAPE}
\label{jetscape}

\begin{figure}[h]
\centering
\includegraphics[width=0.8\linewidth]{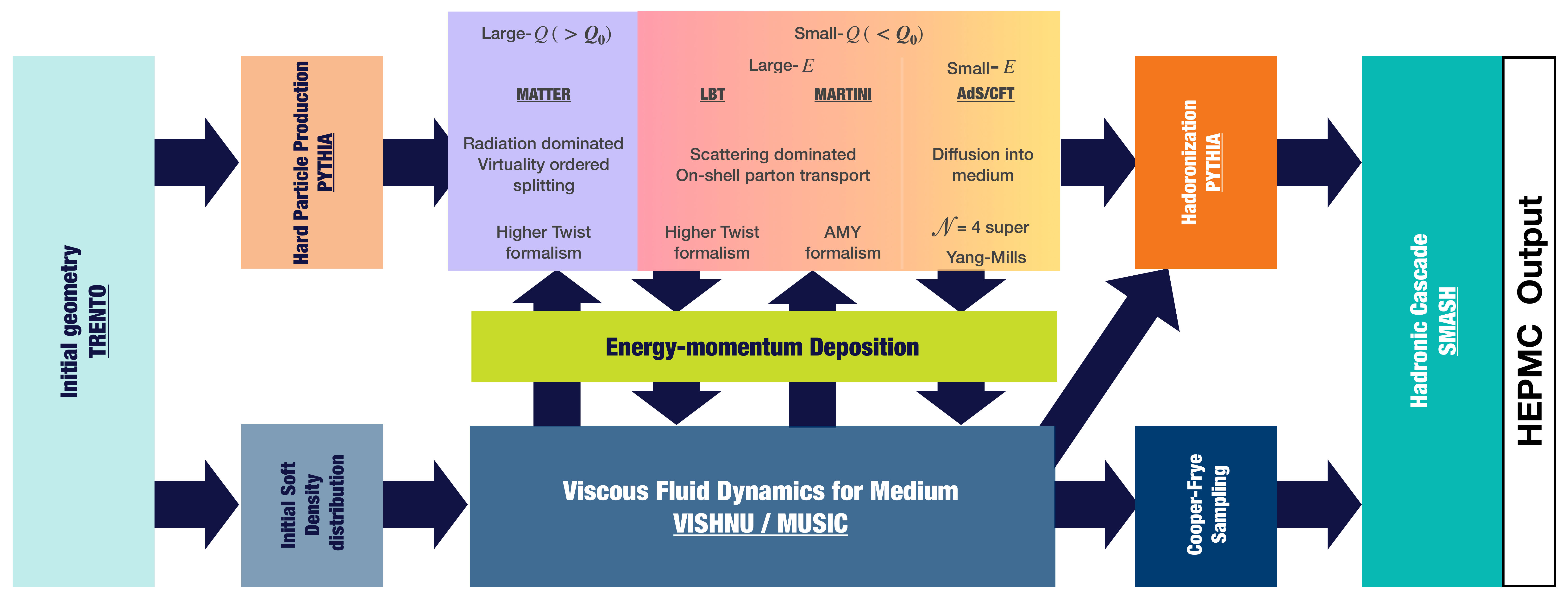}
\caption{Example of Jet-quenching ingredients JETSCAPE Framework}
\label{fig_jetscape}
\end{figure}

The Jet Energy-loss Tomography with a Statistically and Computationally Advanced Program Envelope (JETSCAPE) framework expands upon previous efforts by applying the concept of multi-stage modularity to a virtuality and energy ordered evolution of the jet partons.  As shown in Figure~\ref{fig_jetscape}, the TRENTO model~\cite{Moreland:2015bb} is used to calculate the initial state nuclear geometry.  This model was chosen for its ability to approximate a number of physically motivated distributions as well as interpolate between them.  The hard-scattering cross-sections are calculated with PYTHIA8~\cite{Sjostrand:2014cx}, and evolution of the medium can be calculated with the viscous 2+1D hydrodynamic code, VISHNU~\cite{shen:2016aa}, or the viscous 3+1D codes, MUSIC~\cite{Schenke:2010dia} or CLVisc~\cite{shen:2018aa}.  For the jet evolution, the initial splitting functions can be calculated within MATTER~\cite{Majumder:2013fs}.  Below a virtuality denoted by $Q_0$, the jet evolution may be passed to the Linear Boltzmann Transport (LBT)~\cite{cao:2016aa} or MARTINI~\cite{schenke:2009aa} models.  Initial comparisons in a static medium have shown both LBT and MARTINI to yield consistent results.  A separate module may also be employed at lower virtuality and energy to calculate the parton drag based on the AdS/CFT duality.  The Cooper-Frye formula~\cite{Cooper:1974gr} is used for soft particlization, and hard particle hadronization proceeds through PYTHIA8 routines, modified to provide both a colored or colorless option.  Recent benchmarks with p+p data indicate that the colorless option, developed for the parton-rich heavy-ion environment, may also be preferred for proton collisions~\cite{Kumar:2019um}.  All subsequent soft-hadronic interactions are calculated within SMASH~\cite{weil:2016aa} and final outputs are saved in the HEPMC format.  See~\cite{Putschke:2019tc} for a detailed description of JETSCAPE code and use.

\begin{figure}[h]
\centering
\includegraphics[width=0.8\linewidth]{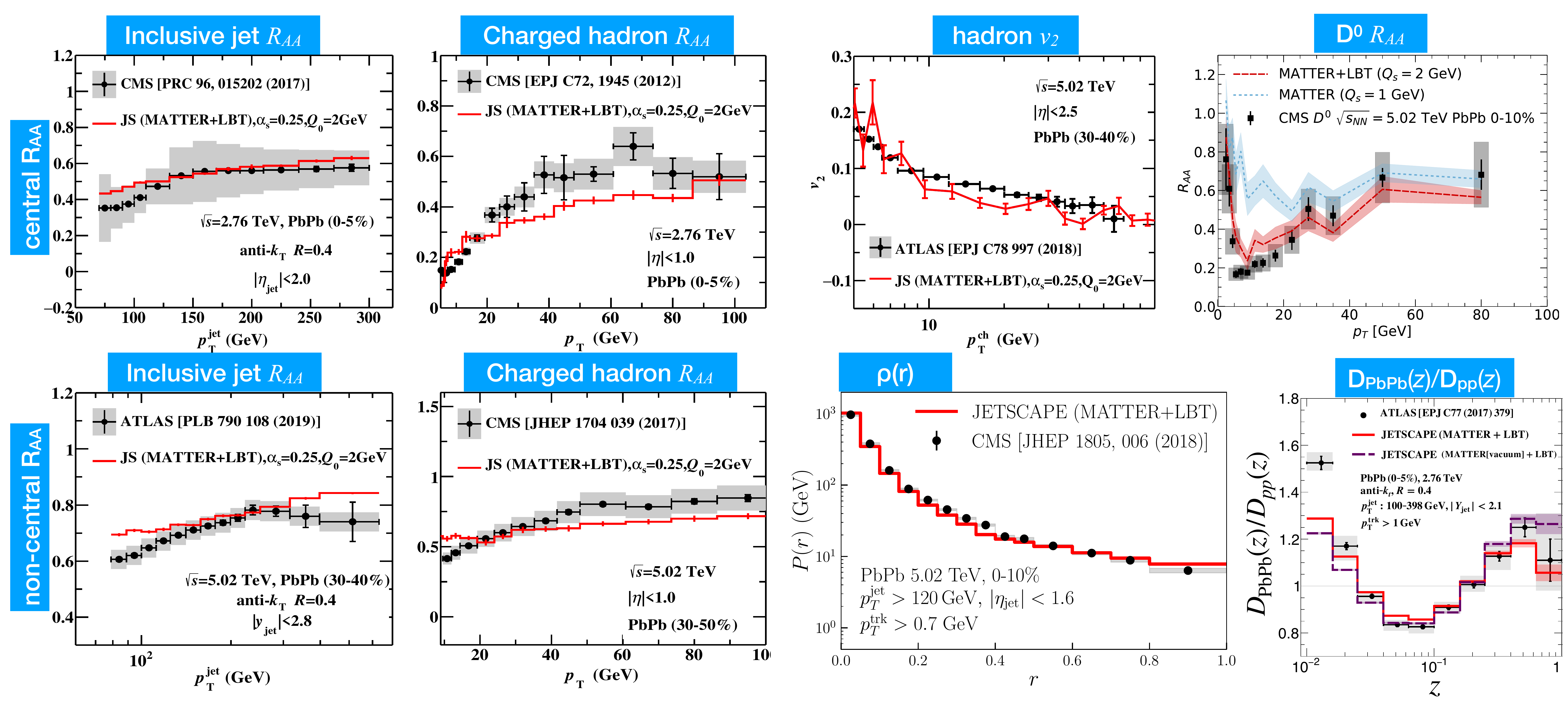}
\caption{Overview of JETSCAPE results presented at Quark Matter 2019.}
\label{fig_jetscape_overview}
\end{figure}

Figure~\ref{fig_jetscape_overview} shows a sample of results shown at Quark Matter 2019 for JETSCAPE run with MATTER+LBT with a virtuality switching parameter $Q_0=2$~GeV.  See the appropriate proceedings~\cite{kumar:2020aa,tachibana:2020aa,vujanovic:2020aa} for detailed discussions of each result.  Two recent results not shown here are a Bayesian statistical analyses of the soft particle production~\cite{paquet:2020aa} and a determination of the virtuality switching parameter, $Q_0$, using temperature dependent parameterization of $\hat{q}/T^3$~\cite{soltz:2019aa}

\section{Framework Usage, Statistics, and Correlated Errors}
\label{errors}

As described earlier, the purpose of a framework is to provide a consistent basis for testing different models and comparing to wide range of experimental observables.  The fact that each model within the framework may have one or more parameters adds another layer of complication to the comparison.  For a limited set of observables and for models with few parameters, a frequentist statistical approach can be used~\cite{Adare:2008cs}, but for a larger numbers of parameters and observables Bayesian approaches are optimal.  To prove this point, note that a chi-squared analysis of soft physics signatures by this author~\cite{Soltz:2012rk} took longer, produced fewer physics insights, and accumulated a smaller number of citations than a concurrent Bayesian analysis by the MADAI Collaboration~\cite{Novak:2013tf}.

If used properly, the Bayesian approach can be used to determine the most likely range of values for input parameters for an assumed set of prior distributions.  The final results should not be overly sensitive to a particular choice of the prior distribution, and the posterior distributions are not meaningful if the model does not give a {\em reasonable description} of all experimental observables.  Models that do not describe the data should eventually be discarded or modified once the underlying reasons for failure are understood, and prior distributions for input parameters should not extend beyond their physically admissible values.  The phrase {\em reasonable description} refers to the probability that a given implementation of the framework describes the data within errors.  The topic of model error is left for future discussion and future conferences, and the treatment of independent statistical and systematic errors is well understood.  However, the proper treatment of correlated errors requires additional information than what is currently being released by most experimental collaborations.

\begin{figure}[h]
\centering
\includegraphics[width=0.8\linewidth]{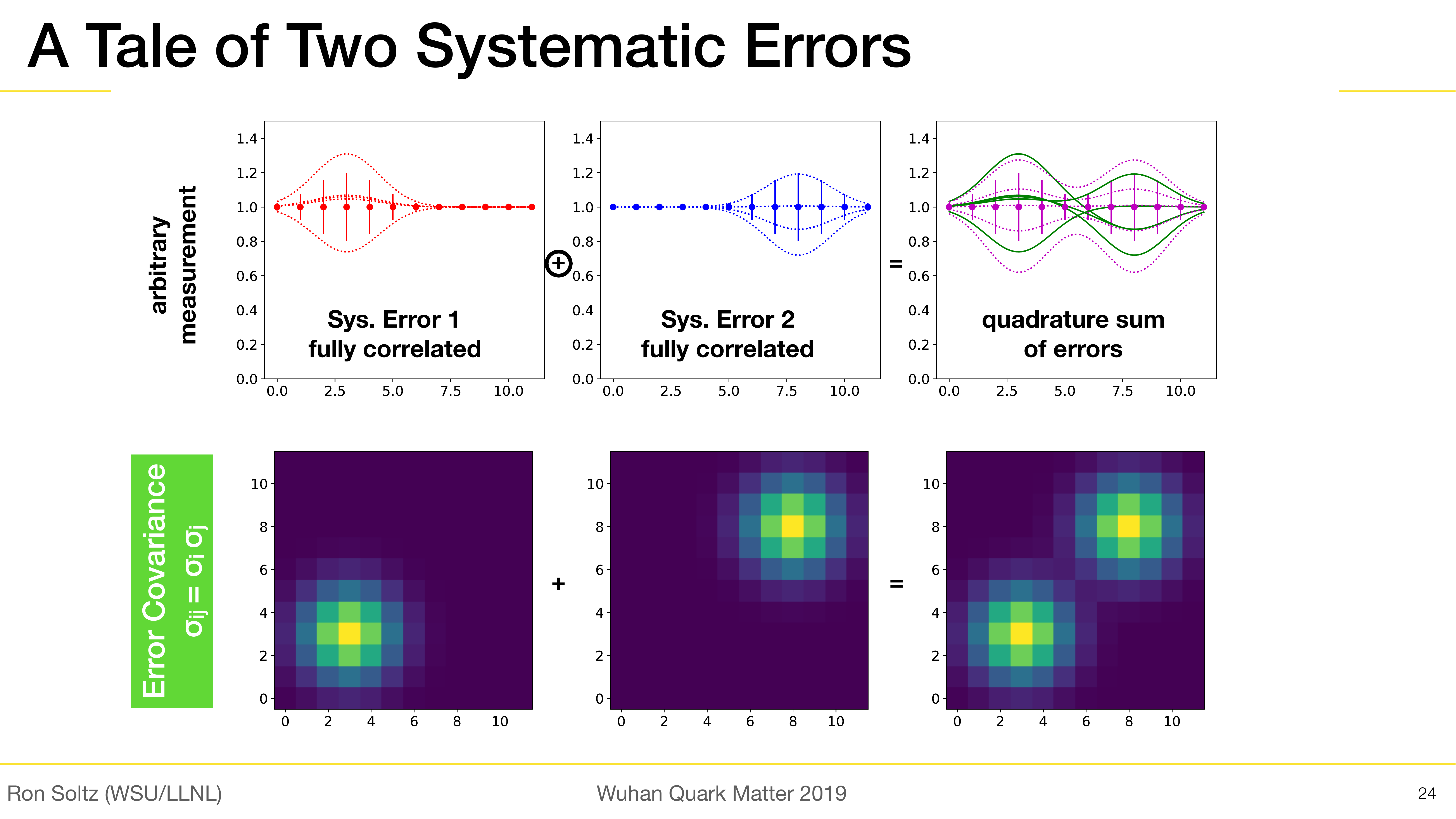}
\caption{Example of two uncorrelated Gaussian-errors and the difference between summing in quadrature (above) vs. summing covariances (below).  The curves are double-Gaussian fits assuming the two sets of errors to be fully correlated (dashed magenta) and fully uncorrelated (solid green)}
\label{fig_err23}
\end{figure}

As an example, consider a hypothetical ratio measurement with two sources of errors, each with a Gaussian profile, at low and high regions of the abscissa.  The errors within each Gaussian region are correlated, but the errors in the separate regions are uncorrelated.  Figure~\ref{fig_err23} shows a characteristic unitary measurement with this error profile.  The top panel shows the data and error band for each region in the left panels, and the quadrature sum of errors on the right.  The lower panel shows the corresponding co-variance error matrices, along with the sum.  
the top right panel shows two fits to the data, the solid-green curves are draws from a distribution that reflects the uncorrelated nature of the two sources of errors, whereas the dashed-magenta curves are drawn from distributions that assume all errors to be fully correlated.  Note that the dashed-magenta curves associated with fully correlated errors do not account for the full range of variation permitted by the data.

\begin{figure}[h]
\centering
\includegraphics[width=0.95\linewidth]{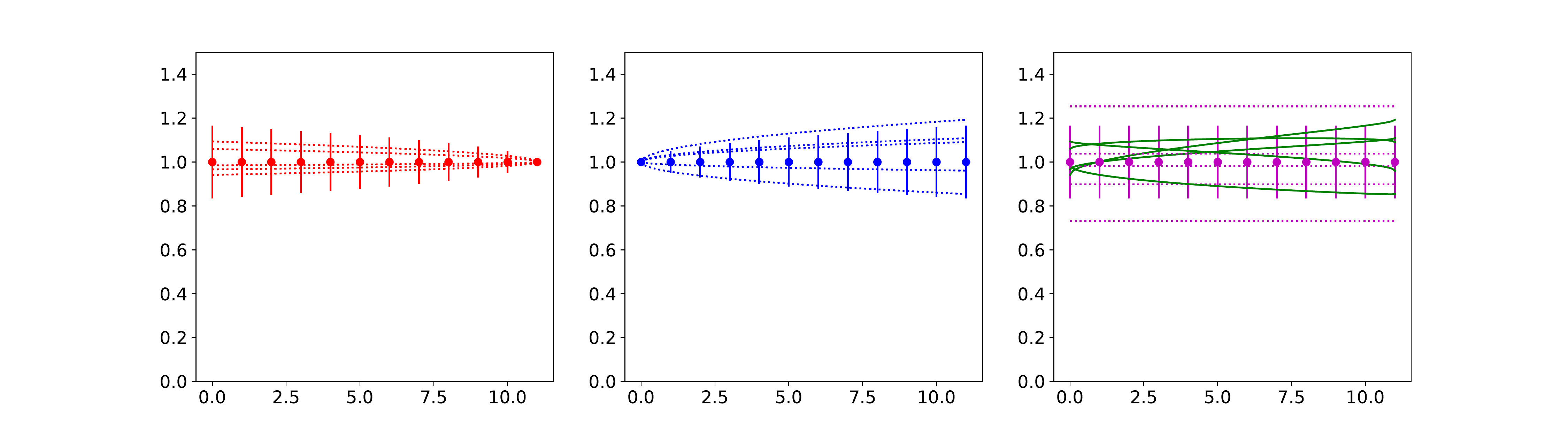}
\caption{Example of two uncorrelated squre-root-errors peaking at low and high values abscissa.}
\label{fig_err4}
\end{figure}

A more realistic example for jet-quenching may be the one shown in Figure~\ref{fig_err4}, which shows two sources of errors that increase in magnitude with the square-root of distance along the abscissa towards the low and high regions, such that the quadrature sum shown on the right displays a constant error band.  The magenta-curves drawn from a fully correlated distributions are all lines of zero slope, whereas the solid-green draws from the true error distributions allow for a monotonic increase or decrease along the abscissa.  To achieve a proper assessment of errors for model comparisons, experimentalists will need to start releasing full covariances for the errors, or at least publish the separate, $p_T$ dependent contributions of errors in cases where the errors are assumed to be fully correlated.  Note that this may also require new methods for plotting data to visualize goodness-of-fit criteria when the systematic error correlations vary significantly with $p_{T}$.  Ultimate, a proper treatment of both experimental and model errors will be needed to realize the full potential of the jet-quenching framework.

\section*{Acknowledgements}
This  work  was  supported  in  part  by  the National Science Foundation within the framework of the JETSCAPE collaboration, Cooperative Agreement ACI-1550300, and the U.S. Department of Energy under contract DE-AC52-07NA27344.


\bibliographystyle{elsarticle-num}
\biboptions{sort&compress}

\end{document}